\documentclass[pdftex,twocolumn,epjc3]{svjour3}

\RequirePackage[T1]{fontenc}

\smartqed

\RequirePackage{graphicx}
\RequirePackage{mathptmx}
\RequirePackage{flushend}
\RequirePackage[numbers,sort&compress]{natbib}
\RequirePackage[colorlinks,citecolor=blue,urlcolor=blue,linkcolor=blue]{hyperref}

\journalname{Eur. Phys. J. C}

\usepackage[vcentermath]{youngtab}
\usepackage{booktabs}
\usepackage{xspace}
\usepackage{amsmath,amssymb}

\newcommand{\ptmiss}{\ensuremath{ p_{\rm{T}}^{\rm{miss}}}\xspace}
\newcommand{\docaiso}{\ensuremath{ \mathrm{DOCA_{ISO}}}\xspace}

\newcommand{\psiDS}{\ensuremath{\psi_{\rm{DS}}}\xspace}

\begin{document}

\title{Prospects on searches for baryonic Dark Matter produced in \boldmath $b$-hadron decays at LHCb}

\author{Alexandre Brea Rodr\'{i}guez \thanksref{e1,addr1}
\and 
Veronika Chobanova \thanksref{e2,addr1}
\and
Xabier Cid Vidal  \thanksref{e3,addr1}
\and
Sa\'{u}l L\'{o}pez Soli\~{n}o \thanksref{e4,addr1}
\and
Diego Mart\'{i}nez Santos  \thanksref{e5,addr1}
\and
Titus Momb\"{a}cher \thanksref{e6,addr1}
\and
Claire Prouv\'e \thanksref{e7,addr1}
\and
Emilio Xos\'{e} Rodr\'{i}guez Fern\'{a}ndez \thanksref{e8,addr1}
\and
Carlos Vázquez Sierra \thanksref{e9,addr2}}

\thankstext{e1}{e-mail: alexandre.brea.rodriguez@cern.ch}
\thankstext{e2}{e-mail: veronika.chobanova@cern.ch}
\thankstext{e3}{e-mail: xabier.cid.vidal@cern.ch}
\thankstext{e4}{e-mail: saul.lopez.solino@cern.ch}
\thankstext{e5}{e-mail: diego.martinez.santos@cern.ch}
\thankstext{e6}{e-mail: titus.mombacher@cern.ch}
\thankstext{e7}{e-mail: claire.prouve@cern.ch}
\thankstext{e8}{e-mail: emilioxose.rodriguez.fernandez@rai.usc.es}
\thankstext{e9}{e-mail: carlos.vazquez@cern.ch}

\institute{Instituto Galego de F\'{i}sica de Altas Enerx\'{i}as (IGFAE), Universidade de Santiago de Compostela, 15782 Santiago de Compostela, Spain  \label{addr1}
\and
 European Organization for Nuclear Research (CERN), Geneva, Switzerland \label{addr2}}

\date{Published: 2 November 2021}

\maketitle

\begin{abstract}
\begin{sloppypar}
A model that can simultaneously explain Dark Matter relic density and the apparent matter-antimatter imbalance of the universe has been recently proposed. The model requires $b$-hadron branching fractions to Dark Matter at the per mille level. The $b$-hadrons decay to a dark sector baryon, \psiDS, which has a mass in the region \mbox{$940 ~\rm{MeV}/c^2\leq m_{\psiDS}\leq 4430~\rm{MeV}/c^2$}. 
In this paper, we discuss the sensitivity of the LHCb experiment to search for this dark baryon, covering different types of topology and giving prospects for Runs 3 and 4 of the LHC, as well as for the proposed Upgrade II. We show that the LHCb experiment can cover the entire mass range of the hypothetical dark baryon.
\end{sloppypar}
\end{abstract}

\sloppy
\section{Introduction}
Despite the long standing success of the Standard Model (SM) of particle physics, there is experimental evidence for New Physics that is not explained by the SM. The most tantalising signs for New Physics are the apparent existence of Dark Matter and the matter-antimatter asymmetry in the universe.
Recently, a model has been proposed that can simultaneously explain these two pressing unknowns~\cite{Elor:2018twp,Alonso-Alvarez:2021qfd}. In this model, neutral $B$ mesons oscillate and then decay into a neutral dark-sector baryon (hereafter called \psiDS) plus SM hadrons. The matter-antimatter imbalance is then proportional~\cite{Elor:2018twp} to
\begin{align*}
\begin{split}
Y_B \propto& \sum_{s,d}A_{SL}^{s,d}\times \mathcal{B}(B^0_{s,d} \rightarrow \psiDS X) = \\ &-\sum_{s,d}\bigg|\frac{ \Delta\Gamma_{s,d}}{\Delta m_{s,d}\cos(\phi_{s,d}-\arg(\Gamma_{12}^{s,d}))}\bigg|\sin(\phi_{s,d}-\arg(\Gamma_{12}^{s,d})) \\ 
&\times \mathcal{B}(B^0_{s,d} \rightarrow \psiDS X)\ ,
\end{split}
\end{align*}
\noindent where the semileptonic asymmetry is defined as 
\begin{equation*}
A^{s,d}_{SL}=\frac{\Gamma(\overline{B}^0_{s,d}\to f)- \Gamma({B}^0_{s,d}\to \bar{f})}{\Gamma(\overline{B}^0_{s,d}\to f)+ \Gamma({B}^0_{s,d}\to \bar{f})}
\end{equation*} 
\noindent with a final state $f(\bar{f})$ that is specific to ${B}^0_{s,d}(\overline{B}^0_{s,d})$.
It has been directly measured by several experiments, being the world averages $A^d_{SL} = (-21 \pm 17)\times10^{-4}$ and \mbox{$A^s_{SL} = (-6 \pm 28)\times10^{-4}$}~\cite{HFLAV18}. 
Alternatively, it can be indirectly inferred through the $B^0_{s,d}$ mixing parameters $\Delta m_{s,d}$, $\Delta\Gamma_{s,d}$, $\phi_{s,d}$ and $\Gamma^{s,d}_{12}$. 
In the SM as well as in New Physics models with negligible contributions to $\Delta F = 1$ penguins, these asymmetries are predicted to be \mbox{$A^d_{SL} = (-4.73\pm0.42)\times10^{-4}$} and $A^s_{SL} = (0.205 \pm 0.018)\times 10^{-4}$~\cite{Lenz:2019lvd}, where similar values are found in Ref.~\cite{Asatrian:2020zxa}. 
Note that, since one needs a positive value for the semileptonic asymmetry in order to obtain a matter-dominated universe, the possibility of having the baryogenesis mechanism lead by $B_s^0$ decays to dark-sector baryons is significantly favoured over those of $B^0$ mesons. 

In order to fulfill the observed matter-antimatter imbalance, the inclusive branching fractions of $b$ hadrons decaying into the dark-sector baryon must be relatively large, $\gtrsim 10^{-4}$, with experimental upper limits
ranging from $\leq10^{-4}$ to $\leq10^{-2}$ depending on the considered \psiDS mass~\cite{Alonso-Alvarez:2021qfd,ALEPH:2000vvi}. Relatively high values could in principle also modify $\Gamma_s/\Gamma_d$ if they have different values for $B_s^0$ mesons compared to $B^0$ mesons\footnote{G. Alonso, private communication.}. 

As for the exclusive branching fractions (as are the examples shown in this paper), these are expected to be at the $\gtrsim 10^{-6}$ level. 
Since the \psiDS baryon must be produced on shell in $B$-meson decays together with at least one proton (to preserve baryon number), its mass cannot be larger than $m_{B_s^0} - m_p$. In addition, to prevent the proton decay, the \psiDS baryon must be heavier than the proton. Hence its mass range is limited to  $940\,\text{MeV}/c^2 \leq m_{\psiDS}\,\leq 4430~\text{MeV}/c^2$.
As an additional remark, we point the reader to Refs.~\cite{Elor:2020tkc,Elahi:2021jia}, which approach the same problem from slightly different angles. 

Due to its unique design and the abundant production of $b$ hadrons in high-energetic $pp$ collisions, the LHCb experiment is ideally suited for searching for the proposed type of baryonic dark matter in $b$-hadron decays~\cite{Chobanova:2020vmx,Borsato:2021aum}. 
Equipped with a highly granular vertex detector very close to the $pp$-interaction point~\cite{LHCbCollaboration:2013bkh}, a precise reconstruction of the $b$-hadron decay vertex can be achieved, which is crucial to obtain a good sensitivity in the search for the dark baryon. Due to baryon-number conservation, searches using $B$ mesons need at least one proton and at least another charged particle to reconstruct the decay vertex. This implies $B$ mesons alone cannot cover the entire phase space of the model. However, LHCb can profit from its large production of $b$ baryons, which do not require the presence of a proton in the final state to preserve baryon number, and which are also heavier than $B_s^0$ mesons.

In this paper, we use as benchmark modes the decays 
$B^0 \rightarrow \psiDS \Lambda(1520) (\rightarrow pK^-)$,
$B^+ \rightarrow \psiDS \Lambda_c(2595)^+ (\rightarrow \pi^+ \pi^- \Lambda_c^+(\rightarrow p K^- \pi^+ ))$,
$\Lambda_b^0 \rightarrow \psiDS K^+\pi^-$,
and
$\Lambda_b^0 \rightarrow \psiDS\pi^+\pi^-$\footnote{Charge-conjugate modes are included unless explicitly stated.} for the search of the dark baryon $\psiDS$ with the LHCb experiment.
These channels are sensitive to different $\mathcal{O}_{q_1 q_2}$ operators out of those introduced in Refs.~\cite{Elor:2018twp,Alonso-Alvarez:2021qfd}. More precisely, $B^0 \rightarrow \psiDS \Lambda(1520)$ decays would be sensitive to $\mathcal{O}_{us}$,  $B^+ \rightarrow \psiDS \Lambda_c(2595)^+$ to $\mathcal{O}_{cd}$, $\Lambda_b^0 \rightarrow \psiDS K^+\pi^-$ to $\mathcal{O}_{us}$, and  $\Lambda_b^0 \rightarrow \psiDS\pi^+\pi^-$ to $\mathcal{O}_{ud}$. Since none of these operators are favored {\it{a priori}} by the mechanism, it becomes essential to probe as many of them as possible.

We explore the possibility of further background suppression by tagging some of these decays via the decay chains \mbox{$\Sigma_b^{(*)\pm}\rightarrow \Lambda_b^0\pi^{\pm}$}.
The projections are based on the detector geometry proposed for the data taking periods Run 3 and Run 4 of the LHCb experiment~\cite{Bediaga:2012uyd} and its Upgrade II~\cite{Aaij:2636441}. The expected sensitivities are computed at corresponding benchmark luminosities of $15$ (Run 3), $50$ \mbox{(Run 4)}, and $300\,\rm{fb}^{-1}$ (Upgrade II)~\cite{Aaij:2636441} of $pp$ collisions at a center-of-mass energy of $14\,\rm{TeV}$.\\
The document is organised as follows:
in Sec.~\ref{sec:gen} we describe the tools used to simulate events and specific decays within the LHCb detector; Section~\ref{sec:sel} discusses potential event selections; and in Sec.~\ref{sec:sens} we describe the estimated sensitivities for the different modes; finally, potential systematic effects are discussed in Sec.~\ref{sec:systematics}; we conclude in Section~\ref{sec:conclusions}.

\section{Event generation and simulation}
\label{sec:gen}
Proton-proton collisions are simulated with \textsc{Pythia 8.226}~\cite{pythia} at a center-of-mass energy of 14 TeV. Note that this does not include $B^+_c$ mesons and therefore backgrounds from $B_c^+$ mesons are not considered in this study.

The four-momenta and origin point of the particles of interest that have been produced by \textsc{Pythia} are stored for further processing. Subsequently, signal $b$ hadrons are decayed according to phase space. No bremsstrahlung is included in the generation of the decays. Background events are obtained from $b\bar{b}$ events in \textsc{Pythia} filtered with selection requirements before proceeding with detector simulation, in order to reduce the amount of events to process.

To obtain the background and signal yields, we use the $b\bar{b}$ cross sections as measured by the LHCb experiment at $\sqrt{s}=13$ TeV and $\sqrt{s}=7$ TeV~\cite{Aaij:2016avz} extrapolated linearly with the center-of-mass energy to $\sqrt{s}=14$ TeV. Hadronisation fractions for $b$ hadrons are obtained from Refs.~\cite{Aaij:2021nyr} and \cite{Aaij:2016avz}, assuming $f_u=f_d$. The production of heavier $b$ baryons and $b$ hadrons with multiple heavy quarks is neglected. We also neglect the effect of pile-up, given the dominant background is composed of single $b\bar{b}$ events and also that we expect the effect of wrong primary vertex association\footnote{i.e. to associate the signal candidate to a wrong primary vertex, which potentially affects the determination of variables such the impact parameter.} to barely affect the sensitivity, given the future vertexing detector of LHCb will be 4D \cite{Aaij:2244311} (including timing), which is known to be very useful to mitigate this effect.
 
The detector is simulated using the code in Ref.~\cite{Chobanova:2020vmx}. The simulated detector elements are the RF foil, the VeloPix (VP) stations, the Upstream Tracker, the Magnet, and the SciFi Tracker~\cite{LHCbCollaboration:2013bkh,LHCbCollaboration:2014tuj}, and are implemented in \texttt{python2}. No calorimetry or particle identification (PID) is simulated at this stage. The particles generated by the procedure described above are passed through the detector elements and yield hits where appropriate. The set of hits is then processed by a track fit algorithm which calculates the track slopes at origin, the momentum, and a point in the early stage of the particle trajectory. Note that through this procedure all detector acceptance effects are accounted for. The simulation neglects occupancy effects and hit inefficiency. To account for this, when obtaining absolute efficiencies, the tracking efficiency is artificially scaled by $98\%$. The detector simulation reproduces at first order the impact parameter and momentum resolutions of the existing full simulation of the LHCb Upgrade, as well as their kinematic dependencies~\cite{LHCbCollaboration:2014tuj, LHCbCollaboration:2013bkh}.

\begin{figure}[tb]
    \centering
    \includegraphics[width=0.5\textwidth]{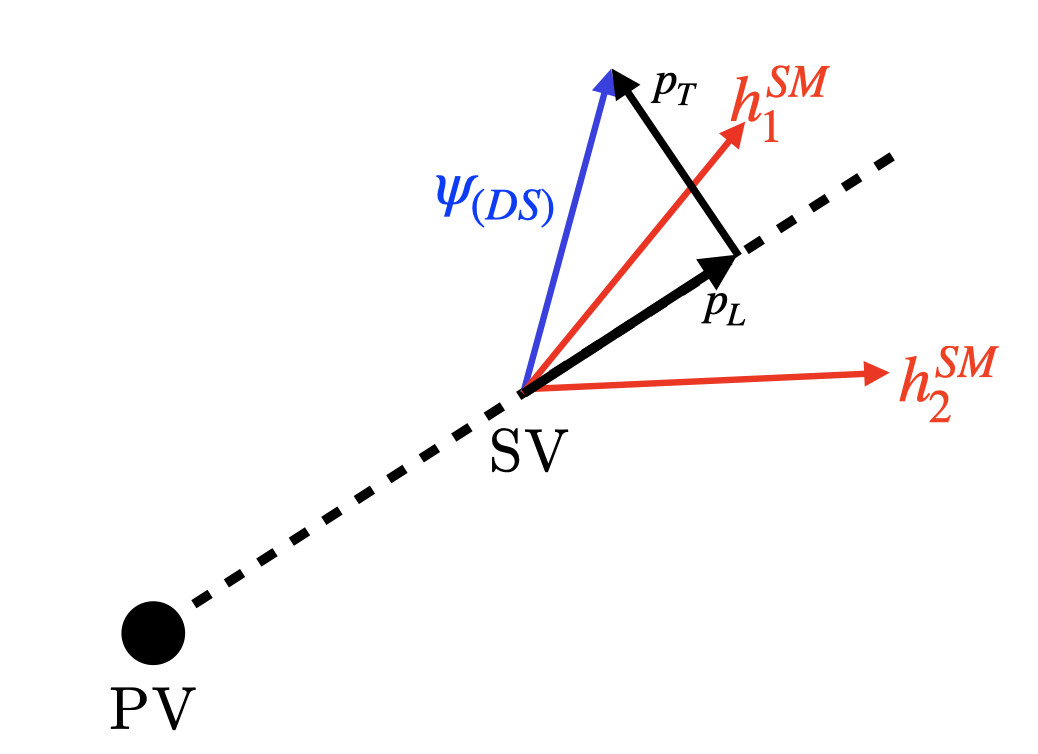}
    \caption{Determination of \ptmiss in $X_b \rightarrow \psiDS h_1^{SM} h_2^{SM}$ decays. For the $B^+\rightarrow \psiDS \Lambda_c(2595)^+$ channel, where more objects are present in the final state, the computation would be analogous. In the figure, PV (SV) represent the primary (secondary) vertices, i.e., the $b$-hadron production (decay) points. }
    \label{fig:ptmiss}
\end{figure}

The final computation of the expected exclusions for the decay channels under evaluation is performed by means of the ``missing'' transverse momentum (\ptmiss). This is defined as the sum of all the momenta of the reconstructed daughters in the direction transverse to the $b$-hadron direction of flight, as illustrated in Fig.~\ref{fig:ptmiss}.
This quantity can only be determined if both the $b$-hadron origin and decay position are known. Both of these properties can be computed experimentally, since they correspond to the $pp$-interaction point, or the origin of all the final-state charged tracks of the $b$ hadron, respectively. The requirement to know the $b$-hadron decay position limits the amount of channels one can study, since this involves knowing the trajectory of the final state SM daughters, which is not possible if these are stable or very long-lived neutral particles. The \ptmiss variable is used in our study to discriminate signal from background. For the $B\rightarrow \psiDS \Lambda$ analyses, the actual expected limit is determined based on this quantity. Furthermore, for the $\Lambda_b^0\rightarrow\psiDS h^+_1 h^-_2$ analyses, also the $h^+_1h^-_2$ invariant mass, $m(h^+_1 h^-_2)$, is very useful since it shows sharp kinematic end points at the \psiDS mass~\cite{Chobanova:2020vmx}, so the limit determination is based on a region selected in the two-dimensional $(m(h^+_1 h^-_2)-\ptmiss)$ plane.

\section{Event selection}
\label{sec:sel}

All tracks involved in the analysis are required to satisfy $p_{\rm{T}}>800~\textrm{MeV}/c$ and $0.1<\textrm{IP}<3$ mm, where $p_{\rm{T}}$ is the transverse momentum in the laboratory frame and IP the impact parameter with respect to the $pp$ interaction vertex\footnote{The lower IP requirement discriminates against particles produced at the $pp$ interaction vertex, while the IP upper requirement helps to remove potential ghost tracks.}. The only exception to this are the pions appearing in the $B^+\rightarrow \psiDS \Lambda_c(2595)^+$ decay for which the $p_{\rm{T}}$ requirement is relaxed to \mbox{250 MeV$/c$}. We assume 100\,\% trigger efficiency.  
The rest of the requirements are channel dependent and are explained in the next subsections.

\subsection{Isolation}

The main background for the targeted decay channels in this analysis is composed of hadrons coming from a $b\bar{b}$ pair. A key feature of this situation is that the background tracks are typically accompanied by other objects, while for the signal the final state is ``isolated'' and no other nearby particles are expected. Based on this feature, we build a quantity that provides discrimination between signal and background. For this, we take all charged particles arising from $b\bar{b}$ pairs (having excluded those forming the signal candidates), and we select only those reconstructed using our fast simulation procedure, with $p_{\rm{T}}>250$ MeV$/c$. With this, and following Refs.~\cite{CidVidal:2019qub,CidVidal:2019urm}, we determine the smallest distance of closest approach (DOCA) to our candidates and use that to measure the isolation. The distribution of this quantity, \docaiso will peak at smaller quantities for background than for signal. The requirement we apply, $\docaiso>0.05$, retains $\sim80\%$ of signal candidates and suppresses $\sim75\%$ of the background.

\subsection{Selection for \boldmath$B\rightarrow \psiDS \Lambda $ decays}

The selection of each $B\rightarrow \psiDS \Lambda$ decay modes is based on the full reconstruction of the $\Lambda$ resonances, applying invariant mass and DOCA requirements to the relevant daughter particles. Furthermore, IP and $p_{\rm T}$ requirements are applied to the $\Lambda$ candidates. The full list of requirements for these two channels can be found in Table~\ref{tab:tight_cuts}, and the main considerations for each of the selections in the next paragraphs.

For the $B^0 \rightarrow \psiDS\Lambda(1520)$ decay mode, the $\Lambda(1520)$ resonance is reconstructed through its decay to a $pK^-$ pair. Each member of the pair is required to be close to each other (applying a DOCA requirement), and an invariant mass requirement to the pair is also applied, so the mother candidate has an invariant mass consistent with that of the $\Lambda(1520)$. Furthermore, the $\Lambda(1520)$ candidate must not be consistent with originating at the $pp$ interaction vertex (IP requirement) and satisfy a minimum $p_{\rm T}$ requirement.

For the $B^+\rightarrow \psiDS \Lambda_c(2595)^+$ channel, two subsequent decays are required to reconstruct the signal candidates, $ \Lambda_c(2595)^+ \rightarrow \Lambda_c^+ \pi^+\pi^-$ and $\Lambda_c^+ \rightarrow pK^-\pi^+$. After applying the track-level requirements described above, the $\Lambda^+_c$ candidates are assumed to be very clean and no further requirements are applied to them. No fake combinations of $\Lambda^+_c$ are considered either, since this is a very clean resonance at LHCb, being narrow in invariant mass, displaced\footnote{In this case there is a ``double'' displacement given these $\Lambda^+_c$ are produced in $B$-meson decays.} and with a final state including a proton, a kaon and a pion, which are separable from each other through PID requirements. Real reconstructed $\Lambda^+_c$ baryons are then combined with pions to form $\Lambda_c(2595)^+$ candidates. Then, as in the $\Lambda(1520)$ case, invariant mass, DOCA, IP and $p_{\rm T}$ requirements are applied to these.

One relevant exclusive background we have considered for both channels is that originating from \mbox{$X_b \rightarrow Y_{\rm SM} \Lambda_c(2595)^+$} and $X_b \rightarrow Y_{\rm SM} \Lambda(1520) $ decays, {\it{i.e.}}, cases in which the baryons accompanying \psiDS are actually produced in $b$-hadron ($X_b$) SM decays together with other SM particles ($Y_{\rm SM}$). Although closer to signal than the combinatorial background, these type of candidates are still separable by means of the isolation and \ptmiss. After all selection requirements, they are seen to contribute $<0.5\%$ compared to the combinatorial background for the $B^0 \rightarrow \psiDS\Lambda(1520)$ channel, so are not considered further in this case. On the contrary, for the $B^+\rightarrow \psiDS \Lambda_c(2595)^+$ channel, they are relevant and become the dominant source of background, so they are accounted for in the sensitivity determination. The main contribution to this type of background in this channel is composed of semileptonic $\Lambda_b$ decays, i.e $\Lambda_b \rightarrow \Lambda_c(2595)^\pm l^\mp \nu_l$, with $l^\mp$ and $\nu_l$ being a charged lepton and corresponding neutrino, respectively.

\begin{table}[h]
\caption{Requirements for each $B\rightarrow \psiDS \Lambda$ decay mode. These values have been chosen in order to maximize the discrimination between signal and background. Here, ``max(DOCA)'' refers to the maximum value of all the DOCA computed for the two-track combinations among the $\Lambda(1520)$ and $\Lambda_c(2595)^+$ daughters, and  $|m_\Lambda-m_{\rm PDG}|$ describes the mass window in the relevant reconstructed invariant $\Lambda$ mass around its known mass \cite{PDG}. 
The IP and $p_{\rm{T}}$ requirements are applied to the $\Lambda(1520)$ and $\Lambda_c(2595)^+$ candidates. 
All the track requirements, introduced in Sec.~\ref{sec:gen}, have also been imposed for every decay mode.}
\label{tab:tight_cuts}
\centering
\resizebox{0.45\textwidth}{!}{
\begin{tabular}{@{}ccc@{}}
\toprule
Variable & $B^+\rightarrow \psiDS \Lambda_c(2595)^+$ & $B^0\rightarrow \psiDS \Lambda(1520)$ \\
 \midrule
 \docaiso  & 0.05 mm & 0.05 mm  \\
 max(DOCA)  & 0.15 mm & 0.075 mm \\
 $|m_\Lambda-m_{\rm PDG}|$ & 20 MeV/$c^2$ & 7.5 MeV/$c^2$ \\
 IP  & 0.07 mm & 0.1 mm \\
 $p_{\rm{T}}$  & 4000 MeV/$c$ & 1000 MeV/$c$ \\
\bottomrule
\end{tabular}}
\end{table}

\subsection{Selection and Multivariate Classifier for \boldmath$\Lambda_b^0\rightarrow \psiDS h_1^+ h_2^-$ decays}
The $\Lambda_b^0\rightarrow \psiDS h_1^+ h_2^-$ decays suffer similarly from a large amount of
combinatorial background from $b\bar{b}\rightarrow h_1^+ h_2^- X$ processes. In order to improve the
discrimination power of the analysis, multivariate classifiers are built using the \ptmiss and the $h_1^+ h_2^-$ reconstructed mass as input variables. Before the training, the following requirements are applied: the two meson tracks must satisfy the requirements introduced at the end of Sec.~\ref{sec:gen}, the DOCA between the two mesons is required to be less than 0.1 mm, the distance of flight in the z direction ($\Delta z$) larger than 5 mm, the transverse momentum of the track larger than 1 GeV/$c$, and the ratio between the impact parameter of the meson pair and the distance of flight in the z direction (IP$(h_1^+ h_2^-)/\Delta z$) is required to be smaller than 0.1. 
The classifier relies on the mathematical method used in Ref.~\cite{Aaij:2011rja} (see also Ref.~\cite{doi:10.1063/1.168691}). A requirement is performed in the multivariate classifier response such that it minimizes the expected limit at a given integrated luminosity and $\psiDS$ mass.

\subsection{Efficiencies}

The efficiencies of the selection requirements proposed in this paper are shown in Table~\ref{tab:effs} for different benchmark masses of the dark baryon, $m_{\psiDS}$.
\begin{table}
\caption{Reconstruction and selection efficiencies for the signal decays described in the text, as obtained from fast simulation, where we added a posteriori a $98\%$ efficiency per track to account for multiplicity effects. Geometry of the detector is considered within $\varepsilon^{REC}$. The efficiencies are shown in \%. The  $\varepsilon^{REC\&PT}$ efficiency implies that all tracks have been required to have a transverse momentum greater than $800~ \text{MeV}/c$, except for $B^+\rightarrow \psiDS \Lambda_c(2595)^+$ decays, where this requirement is relaxed to $250~ \text{MeV}/c$ for pions. The $\varepsilon^{SEL/REC}$ efficiency includes multivariate and isolation requirements when applicable. $\varepsilon^{PID/SEL}$ refers to the efficiency assumed for the PID requirements, as explained in the text. Note this does not apply to modes with no protons or kaons in the final state. Finally, $\varepsilon^{TOTAL}$ is the product of $\varepsilon^{REC}$, $\varepsilon^{SEL/REC}$  and $\varepsilon^{PID/SEL}$. The number next to \psiDS refers to $m_{\psiDS}$, given in MeV$/c^2$.}
\label{tab:effs}
\vspace{.2cm}
\resizebox{0.5\textwidth}{!}{
\begin{tabular}{c c c  c  c c} 
\toprule
 Decay mode    &  $\varepsilon^{REC}$ &  $\varepsilon^{REC\&PT}$ & $\varepsilon^{SEL/REC}$ & $\varepsilon^{PID/SEL}$ & $\varepsilon^{TOTAL}$\\
\midrule
$\Lambda_b^0\rightarrow\psiDS(940) K^+\pi^-$     & 7.6 & 5.1 & 13.2 & 74.2     & 0.74\\
$\Lambda_b^0\rightarrow\psiDS(940) \pi^+\pi^-$     & 7.3 & 4.8  & 14.3 & -     & 1.04\\
$\Lambda_b^0\rightarrow\psiDS(1500) K^+\pi^-$     & 7.7 & 4.8 & 10.8 & 76.0    & 0.63\\
$\Lambda_b^0\rightarrow\psiDS(1500)  \pi^+\pi^-$     & 7.3 & 4.5  & 12.5 & -   & 0.91\\
$\Lambda_b^0\rightarrow\psiDS(2000)  K^+\pi^-$     & 7.7 & 4.5 & 6.67 & 79.2   & 0.41\\
$\Lambda_b^0\rightarrow\psiDS(2000)  \pi^+\pi^-$     & 7.4 & 4.1  & 10.9 & -   & 0.81\\
$\Lambda_b^0\rightarrow\psiDS(2400)  K^+\pi^-$     & 7.8 & 4.1 & 9.11 & 80.8   & 0.57\\
$\Lambda_b^0\rightarrow\psiDS(2400)  \pi^+\pi^-$     & 7.4 & 3.7  & 8.89 & -   & 0.66\\
$\Lambda_b^0\rightarrow\psiDS(4340)  K^+\pi^-$     & 8.1 & 1.2 & 3.12 & 88.0   & 0.22\\
$\Lambda_b^0\rightarrow\psiDS(4340)  \pi^+\pi^-$     & 7.4 & 0.97  & 2.88& -   & 0.21\\
$\Lambda_b^0\rightarrow\psiDS(4470)  K^+\pi^-$     & 8.2 & 0.91 & 2.18 &  87.7 & 0.16\\
$\Lambda_b^0\rightarrow\psiDS(4470)  \pi^+\pi^-$     & 7.4 & 0.79  & 2.23 & -  & 0.17\\
$B^+\rightarrow \psiDS(940) \Lambda_c(2595)^+  $  & 5.7 & 1.3  & 20.9 & 56.7   & 0.68\\
$B^+\rightarrow \psiDS(1500) \Lambda_c(2595)^+ $ & 5.6 & 1.1  & 19.3 & 56.6    & 0.61\\
$B^+\rightarrow \psiDS(2000) \Lambda_c(2595)^+ $  & 5.3 & 0.88  & 18.5 & 57.2  & 0.56\\
$B^+\rightarrow \psiDS(2400) \Lambda_c(2595)^+ $ & 4.9 & 0.66  & 16.4 & 57.1   & 0.46\\
$B^0 \rightarrow \psiDS (940) \Lambda(1520) $   & 13.3 & 12.7 & 38.1 & 56.9    & 2.88\\
$B^0 \rightarrow  \psiDS (1500) \Lambda(1520) $ & 13.7 & 12.9 & 36.6 &  56.4   & 2.83\\
$B^0 \rightarrow  \psiDS (2000) \Lambda(1520) $ & 13.5 & 12.6 & 35.5 & 56.8    & 2.72\\
$B^0 \rightarrow \psiDS(2400)  \Lambda(1520) $ & 13.3 & 6.8 & 34.5 & 56.6      & 2.6\\
$B^0 \rightarrow \psiDS(3500) \Lambda(1520) $  & 12.1 & 3.5 & 23.6 & 56.5      & 1.61\\
\bottomrule
\end{tabular}}
\end{table}
We also simulate the effect of applying PID requirements that would make the background contribution from misidentified particles negligible. For kaons, we take the expected kaon identification efficiency for a Delta Log Likelihood (DLL) $\rm{DLL}_{K\pi}>5$ requirement as a function of momentum \cite{Collaboration:1624074}, and apply it to our samples, which provides a penalty efficiency factor. For protons we follow a similar procedure, applying $\rm{DLL}>5$ requirements to discriminate protons against kaons and pions. In this case, since no projections exist for the LHCb upgrades, we take the measured numbers from the data-taking periods Run 1 and Run 2 of the LHCb experiment~\cite{Adinolfi:2012qfa}. 

\subsection{Tagging}
\label{sec:tagging}
\begin{figure}[tb]
    \centering
    \includegraphics[width = 0.46\textwidth] {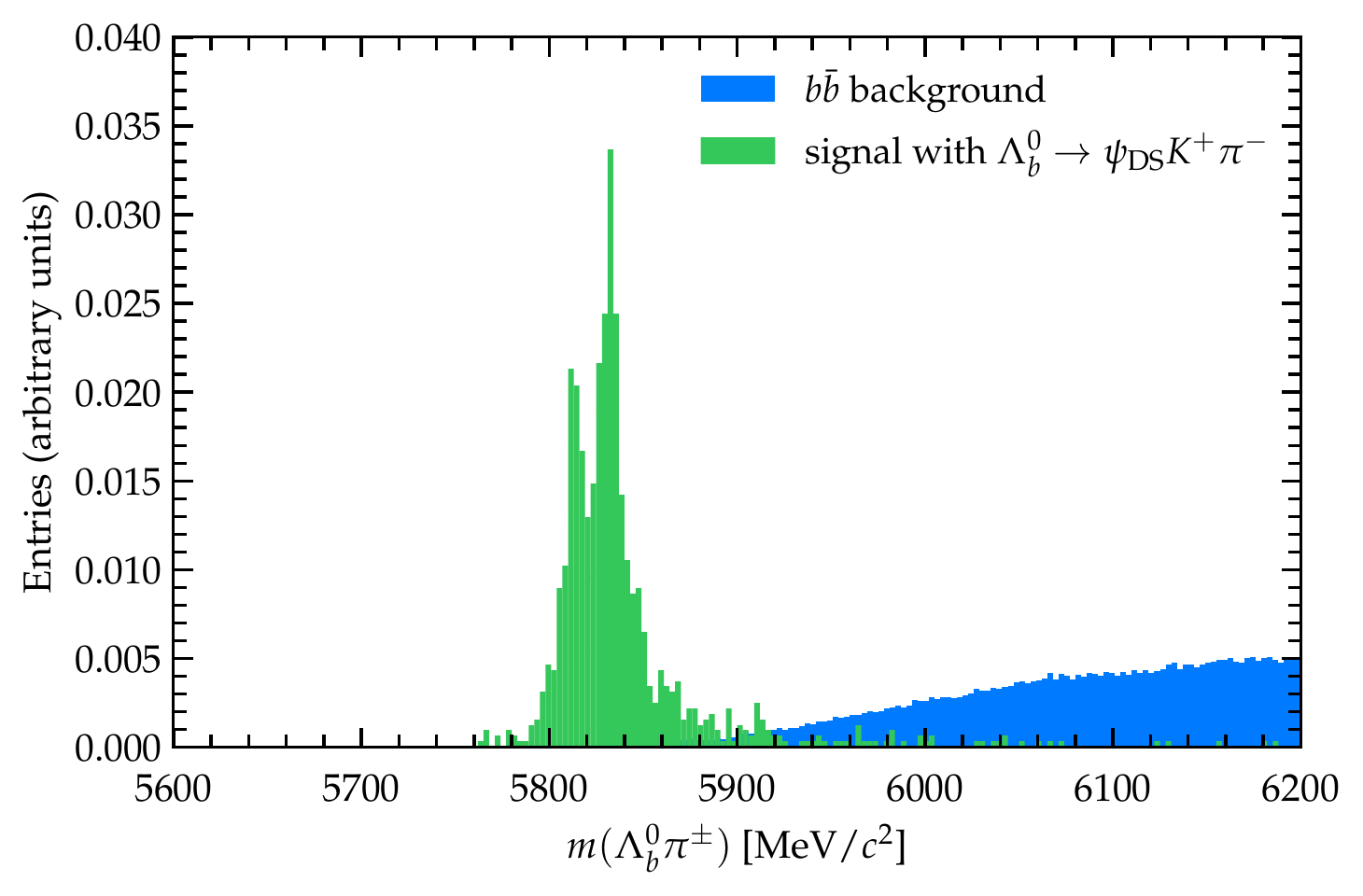}
    \hfill
    \includegraphics[width = 0.46\textwidth]{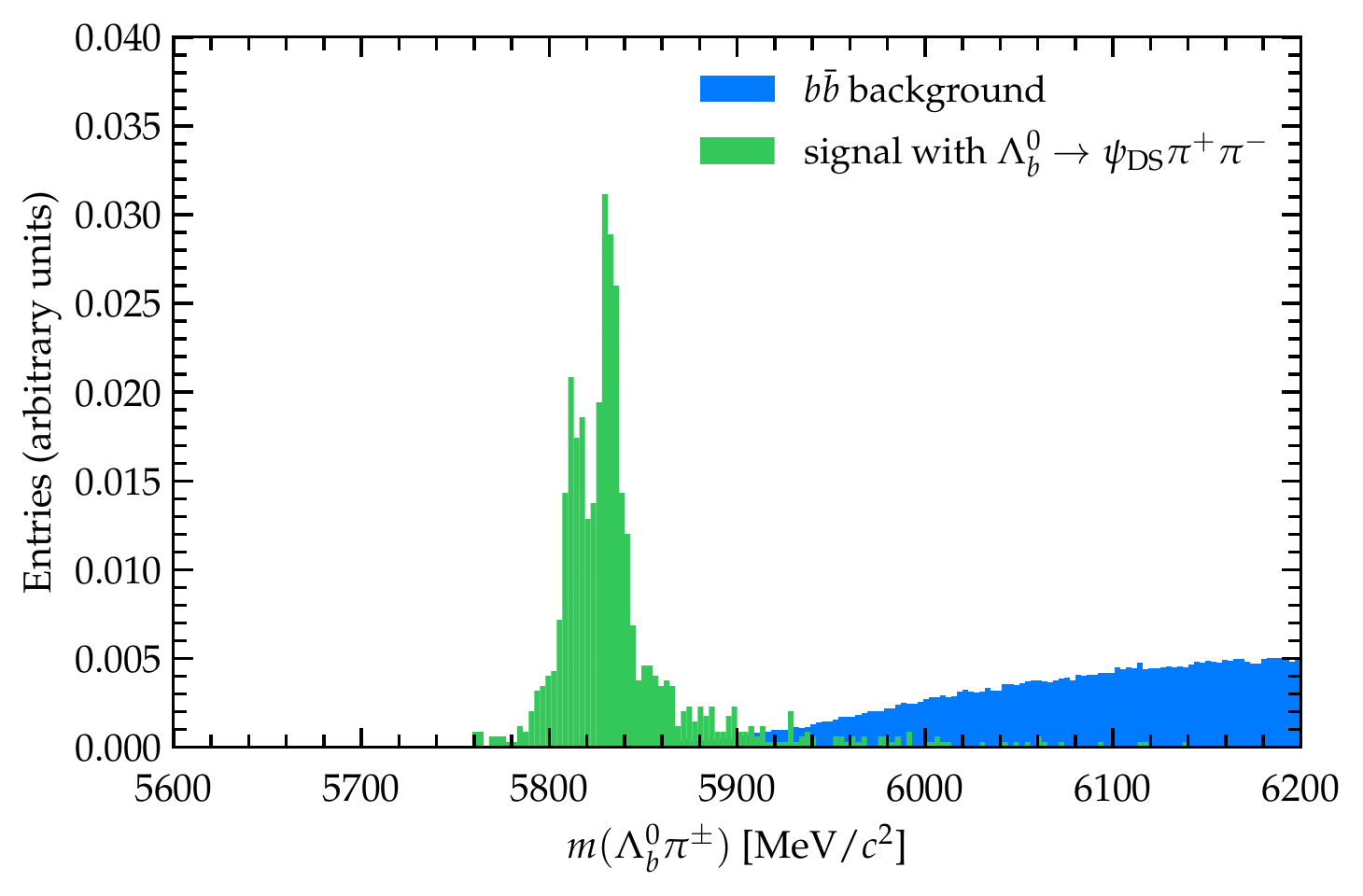}
    \caption{Normalised distributions of the reconstructed invariant mass of $\Sigma_b^{(*)\pm}\rightarrow\Lambda_b^0(\rightarrow\psiDS h^+_1h^-_2)\pi^\pm$ candidates in the simulated samples. The mass of the $\Lambda_b$ baryon is a fixed parameter in this case. Hatched green: signal, blue: $b\bar{b}$ background. Top: $\Lambda_b^0\rightarrow\psiDS K^+\pi^-$ candidates. Bottom: $\Lambda_b^0\rightarrow\psiDS \pi^+\pi^-$ candidates. The tagging of $\Lambda_b^0$ allows a clear distinction between signal and background samples.}
    \label{fig:tagging}
\end{figure}

A useful way to deal with background in $\Lambda_b^0$ decays while getting additional kinematic information, is to select those $\Lambda_b^0$ particles coming from $\Sigma_b^{(*)\pm}$ as suggested in Ref.~\cite{Stone:2014mza}. About one third of $\Lambda_b^0$ baryons are coming from $\Sigma_b^{(*)\pm}$ according to \textsc{Pythia} \cite{pythia}. More importantly, from Ref.~\cite{Aaij:2018tnn} one can infer that LHCb yields approximately one tagged $\Lambda_b^0$ per 10 untagged $\Lambda_b^0$, including the effect of reconstructing and selecting the associated slow pion from the decay $\Sigma_b^{(*)\pm}\rightarrow\Lambda_b^0\pi^{\pm}$. 
Most background events will have plenty of slow prompt pions coming from the same collision point as the $\Lambda_b^0$ candidate. However, background events are not expected to peak at the $\Sigma_b^{(*)}$ mass, while signal events do (see Fig.~\ref{fig:tagging}). This could allow extra selection requirements for further background reduction, as well as a good signal confirmation due to the distinctive two-peak pattern. 
In these sensitivity studies the usage of tagging is not included, though.

As for the $B\rightarrow \psiDS \Lambda$ modes, tagging would also be possible in principle for the $B^+\rightarrow \psiDS \Lambda_c(2595)^+ $ channel through $B^+$ mesons produced in $B_{s2}^{\ast 0}\rightarrow B^+ K^-$ decays. Around 20\% of all $B^+$ mesons are produced in a $B_{s2}^{\ast 0}$ decay, according to \textsc{Pythia}. Note that this type of tagging has already been used at LHCb to search for the lepton flavor violating $B^+ \rightarrow K^+ \mu^- \tau^+$ decay \cite{Aaij:2020mqb}, where the $\tau$ lepton was effectively treated as a missing particle. This search achieved upper limits on the branching fraction of this decay at the level of $10^{-5}$, using an LHCb data set accounting for an integrated luminosity of 9 fb$^{-1}$. 

\section{Estimated statistical sensitivities}
\label{sec:sens}

\begin{figure}[tb]
    \centering
    \includegraphics[scale = 0.5] {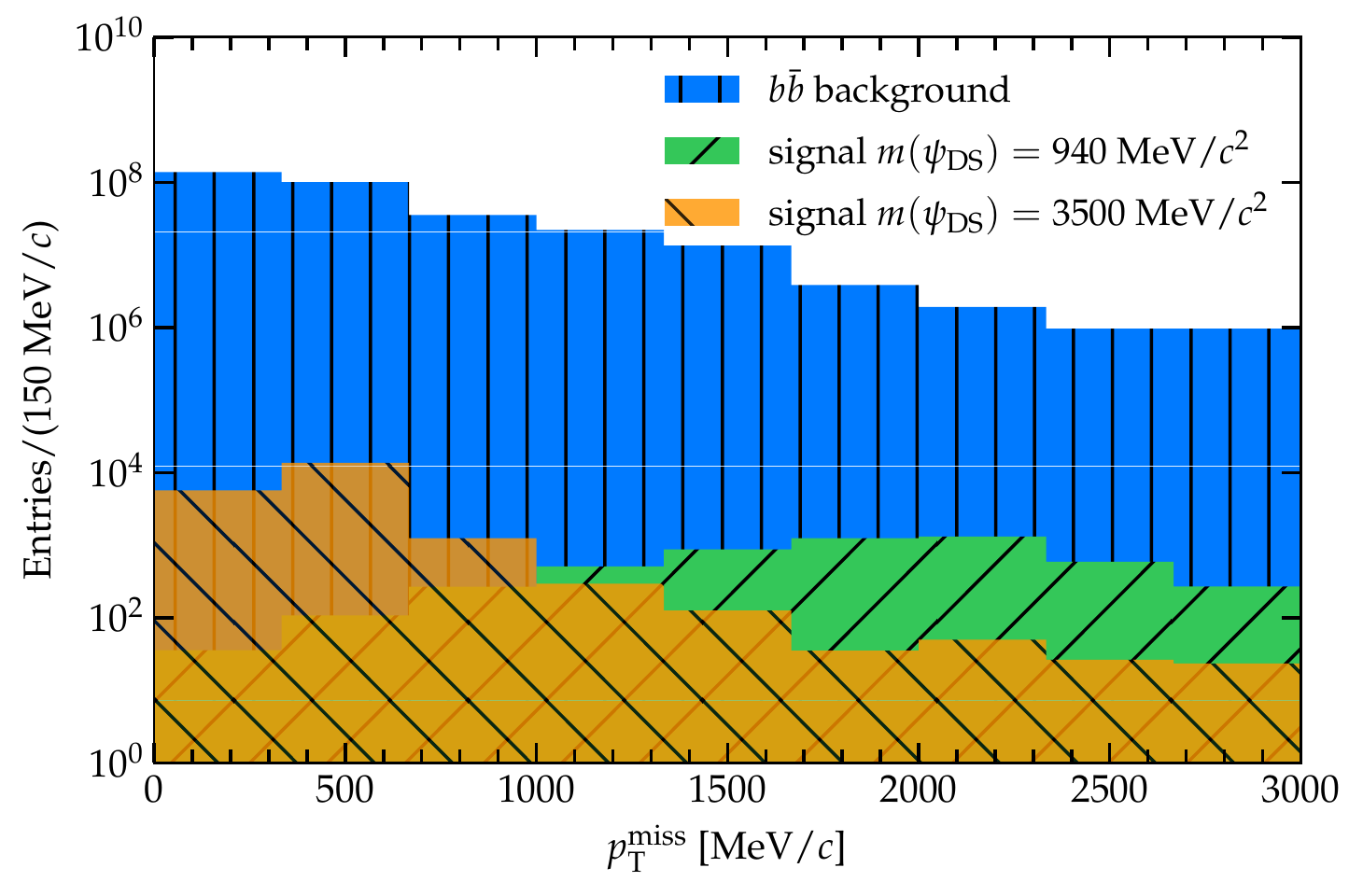}
    \includegraphics[scale = 0.5]{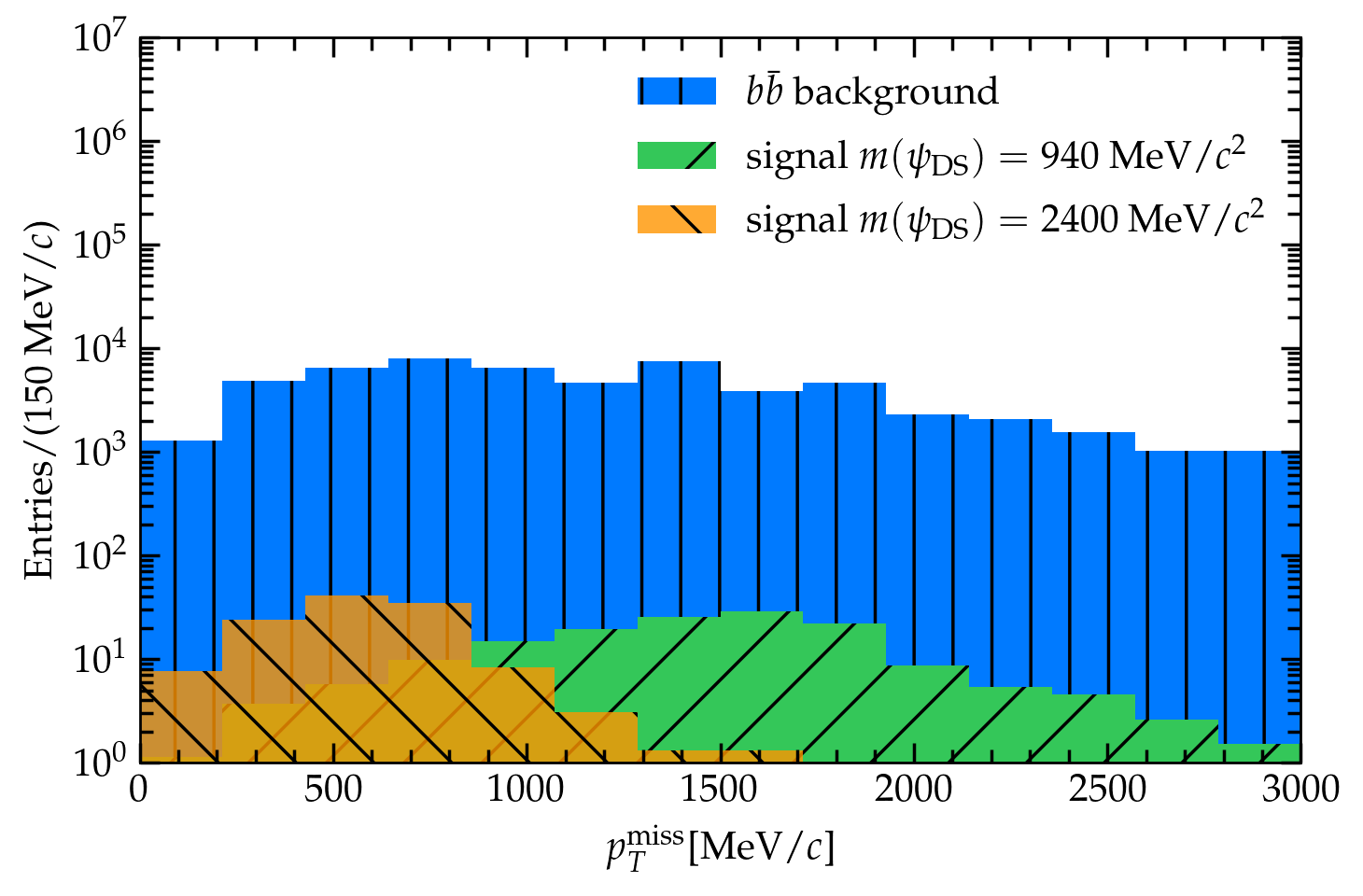}
    \caption{Distributions of \ptmiss for $B\rightarrow\psiDS\Lambda$ and associated backgrounds. Top: $B^0 \rightarrow  \psiDS \Lambda(1520) $. Bottom: $B^+\rightarrow \psiDS \Lambda_c(2595)^+$. Blue: background from inclusive $b\bar{b}$ events. Orange: signal for a high mass (close to phase space limit) dark baryon. Green: signal for a dark baryon at 940 MeV$/c^2$. The signal and background yields correspond to those expected at LHCb at 15 fb$^{-1}$, with the signal branching fractions assumed to be those that could be excluded at 95\% CL (see Sec.~\ref{sec:sens}). No systematic uncertainties are assumed in these plots. }
    \label{fig:mpt}
\end{figure}

\begin{figure}[tb]
    \centering
    \includegraphics[scale = 0.48] {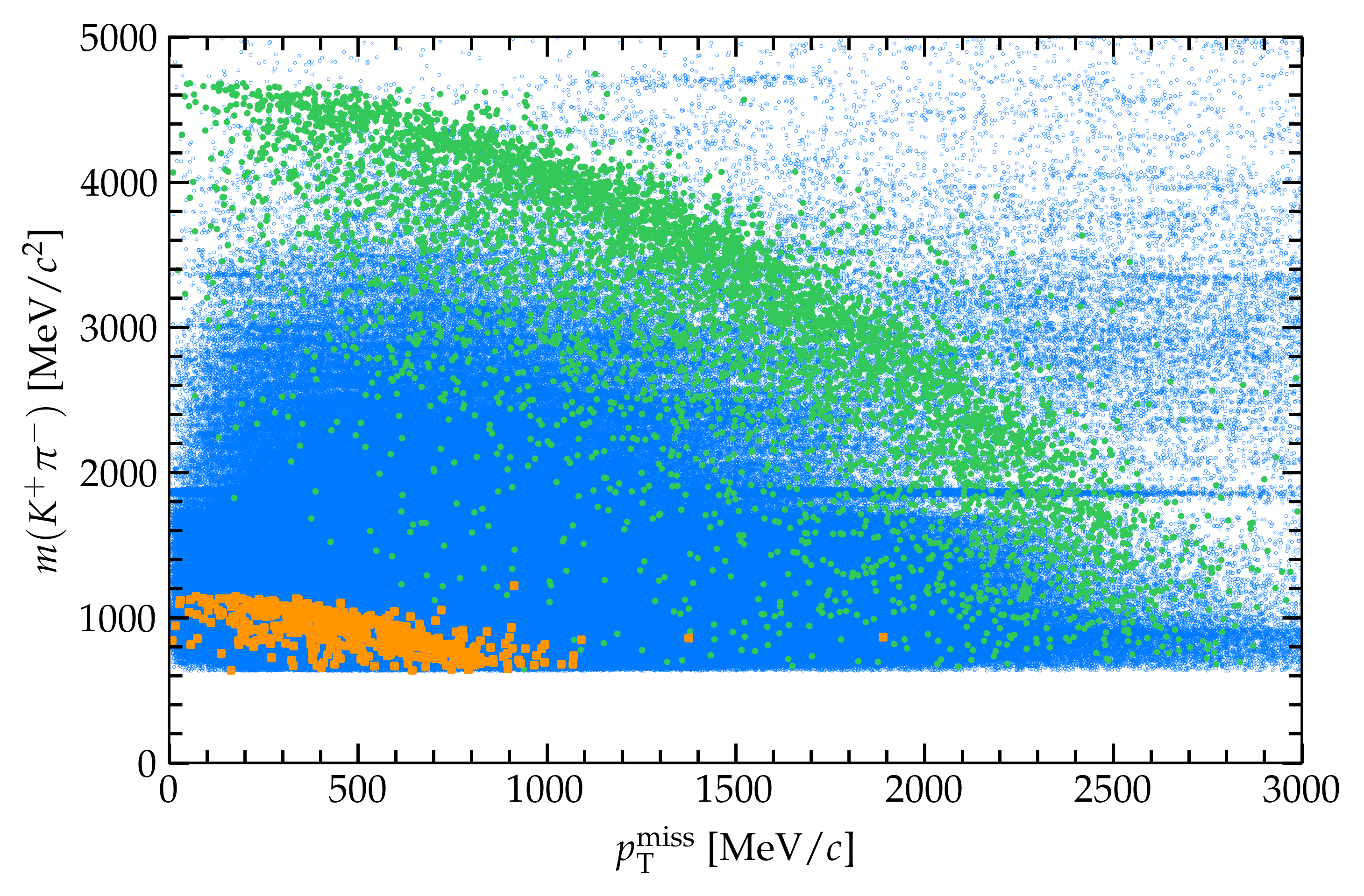}
    \includegraphics[scale = 0.48]{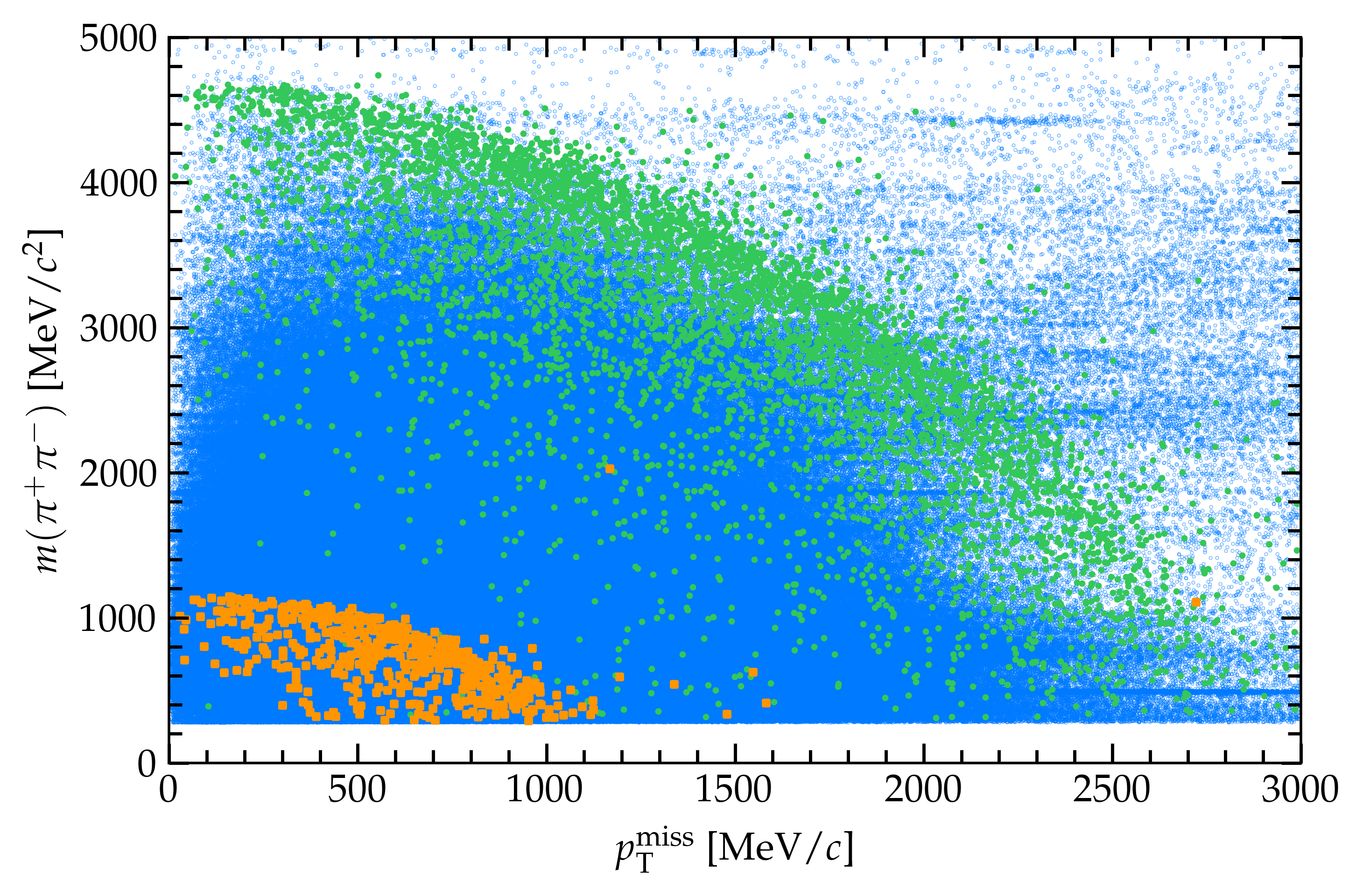}

    \caption{Reconstructed invariant mass versus \ptmiss for $\Lambda_b^0\rightarrow \psiDS h^+_1 h^-_2$ candidates. Top: $\Lambda_b^0\rightarrow \psiDS K^+\pi^-$. Bottom: $\Lambda_b^0\rightarrow \psiDS \pi^+\pi^-$. Blue dots: background from inclusive $b\bar{b}$ events, green circles: signal for a dark baryon at 940 MeV$/c^2$, orange squares: signal for a dark baryon at 4470 MeV$/c^2$. Background from light mass resonances, such as $K^*(892)\rightarrow K^+\pi^-$, $D^0\rightarrow K^+\pi^-$, and $K_S^0\rightarrow \pi^+\pi^-$ is clearly visible in the plots. No systematic uncertainties are assumed in these plots.}
    \label{fig:scatter}
\end{figure}

As discussed in Sec.~\ref{sec:gen}, we obtain the sensitivities from the distributions of \ptmiss, or \ptmiss versus $m(h^+_1h^-_2)$ for signal and background candidates surviving the selections introduced in Sec.~\ref{sec:sel}. 
In order to maximize our sensitivity, the number of observed events in background of the $B$-meson decay modes are obtained in bins of \ptmiss, while for the $\Lambda_b^0$ decay modes a single two-dimensional region as a function of \ptmiss and $m(h^+_1h^-_2)$ is considered.
Figures~\ref{fig:mpt} and~\ref{fig:scatter} show the background and signal distributions (for different values of the dark baryon mass) for the $B\rightarrow \psiDS \Lambda$ and $\Lambda_b^0 \rightarrow \psiDS h_1 h_2$ decay modes, respectively. 

Exclusion limits on the decay branching fractions of the four $B$ and $\Lambda_b^0$ decay modes are then computed using these distributions, with the {\it{Modified Frequentist confidence level}} (CL$_s$) method, described in Refs.~\cite{CLs,CLsv2}. These values are obtained per decay mode and benchmark $\psiDS$ mass hypotheses as listed in Table~\ref{tab:effs}. 

\begin{figure}[tb]
    \centering
    \includegraphics[width = 0.4\textwidth] {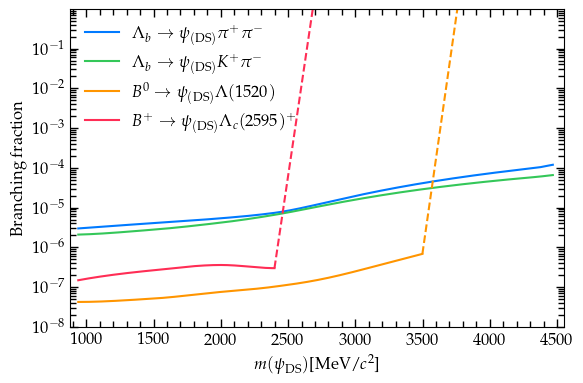}
    \includegraphics[width = 0.4\textwidth] {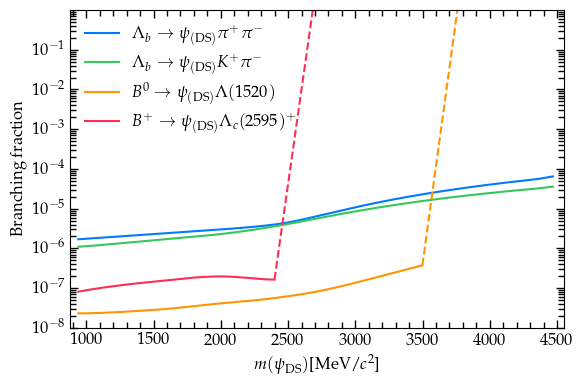}\hfill
    \includegraphics[width = 0.4\textwidth] {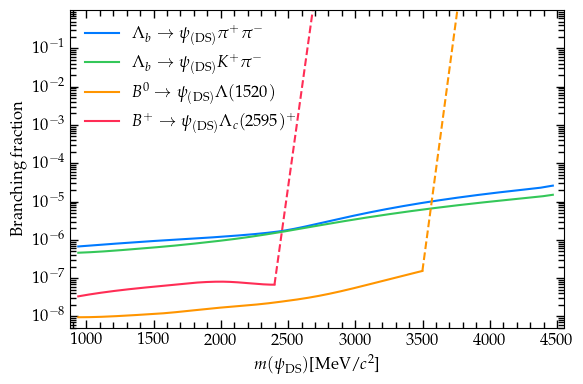}
    \caption{Projected statistical sensitivity in terms of branching fractions excluded at 95\% CL at, from top to bottom, $15~\textrm{fb}^{-1}$, $50~\textrm{fb}^{-1}$ and $300~\textrm{fb}^{-1}$. All three curves were obtained by interpolation, having used an exponential extrapolation (dashed line) in the abrupt change of slope for the $B$ decays in order to account for the detector's inability to reconstruct neither the $\Lambda(1520)$ nor the $\Lambda_c(2595)^+$ baryons at the limit of the phase space, i.e., whenever $ m_{\psiDS}$ reaches the mass difference of the SM mother meson and daughter baryon.}
    \label{fig:sens_15_50}
\end{figure}

Finally, we consider the limits at 95\% C.L. per $\psiDS$ mass hypothesis and we interpolate these values to cover intermediate mass points, as presented in Fig.~\ref{fig:sens_15_50}. For $B$-meson decay modes an exponential extrapolation is used for masses above 3.5(2.5) GeV/c$^2$, to account for the inability to reconstruct the $\Lambda$ mesons due to the limited phase space.

For all the decay modes, branching fractions down to \mbox{$(1$ -- $5){\times}10^{-6}$} for a dark baryon mass between 1 and \mbox{2.5 GeV/c$^2$}, could be excluded. 
The best limits are achieved in the low $m_{\psiDS}$ region and are given by the $B^0$ mode we used, because of the complex final state. High masses are only accessible with $\Lambda_b^0$ decays. It should be noted that at high masses the differences between the inclusive and exclusive branching fraction are smaller, and hence higher branching fractions are expected.

\section{Systematic uncertainties}
\label{sec:systematics}


It should be noted that in these analyses, the background yields are typically larger than in other searches, such as rare decays. Hence, systematic uncertainties related to background yield or modelling can potentially limit the search. For instance, with a background yield systematic uncertainty of $1\%$ all the searches shown in this study would be already systematically limited with $1\,\rm{fb}^{-1}$, and with a systematic uncertainty of 1 per mile, they would be systematically limited after  $\sim 15\,\rm{fb}^{-1}$. Hence, further background suppression, such as usage of the $\Sigma_b^{(*)}$ (or $B^{\ast 0}_{s2}$) tagging, more complex decay chains or the use of control regions might be needed to perform the proposed searches at significantly higher integrated luminosity. The presence of systematic uncertainties might also risk reaching the whole region of theoretical interest, with branching fractions $\gtrsim 10^{-5}$ for the channels of interest and any value of $ m_{\psiDS}$. Background modeling will of course be particularly crucial when $m_{\psiDS}$ approaches the mass of an existing SM resonance, since isolation requirements only reduce the backgrounds from $b\bar{b}$ pairs, but do not grant that the remaining signal candidates are not coming from exclusive backgrounds. More details about the effect of systematic uncertainties are explained in \ref{app:systematics}.

\section{Conclusions}
\label{sec:conclusions}
The B-mesogenesis mechanism predicts branching fractions of $b$-hadron decays to dark baryons at relatively large values ($\sim 10^{-2}-10^{-6}$). The closer the CP-violation sources are to the SM prediction, the higher those branching fractions need to be in order to fulfill the baryon asymmetry of the universe.
In our study, we show that LHCb Upgrade will have the statistical sensitivity to search for branching fractions at the $10^{-5}$ level or better for the entire mass range of the dark baryon \psiDS. Multi-body final states, such as $B^+\rightarrow \psiDS(940) \Lambda_c(2595)^+$, show the best performance at lower values of $m_{\psiDS}$ while two-track final states are needed to reach higher masses, with $\Lambda_b^0$ decays allowing to test the full allowed mass range. Systematic uncertainties could play an important role in the sensitivity, and hence very precise background modelling and suppression could be needed. 

Together with a very precise measurement of the $B_s^0$ mixing parameters and improved theoretical understanding of the model, the LHCb upgraded experiment could arguably exclude completely, or confirm, B-mesogenesis as the underlying mechanism for the baryon asymmetry of the universe as well as the explanation for the Dark Matter problem.

\begin{acknowledgements}
We would like to thank Miguel Escudero, Gonzalo Alonso-\'Alvarez, Gilly Elor, Luca Silvestrini, Kristof De Bruyn, Lucia Grillo, and Alex Lenz for useful discussions.
This work has received financial support from Xunta de Galicia (Centro singular de investigaci\'on de Galicia accreditation 2019-2022), by European Union ERDF, and by  the “Mar\'ia  de Maeztu”  Units  of  Excellence program  MDM-2016-0692  and  the Spanish Research State Agency. In particular: the work of X.C.V. is supported by MINECO (Spain) through the Ram\'{o}n y Cajal program RYC-2016-20073 and by XuntaGAL under the ED431F 2018/01 project.
\end{acknowledgements}

\appendix
\section{Effect of systematic uncertainties}
\label{app:systematics}
The sensitivity results presented in this document make no assumptions on the systematic uncertainties. However, as mentioned in Sec.~\ref{sec:systematics}, one would expect them to limit the actual experimental analysis. We expect the dominant systematic uncertainty to arise from the knowledge of the actual background yield, which originates, among others, from the uncertainty in the $b\bar{b}$ cross section \cite{Aaij:2016avz}. While any experimental analysis will do its best to {\it cure} this uncertainty, this will eventually limit the precision to be achieved. Potential strategies to mitigate this uncertainty include the use of heavier $b$ resonances to tag the events of interest, as discussed in Sec.~\ref{sec:tagging}, the introduction of smarter selection criteria to discriminate the signal against the background or the use of control regions that would provide a handle to determine the actual background content in the signal region.

\begin{figure}[tb]
    \centering
    \includegraphics[width = 0.5\textwidth] {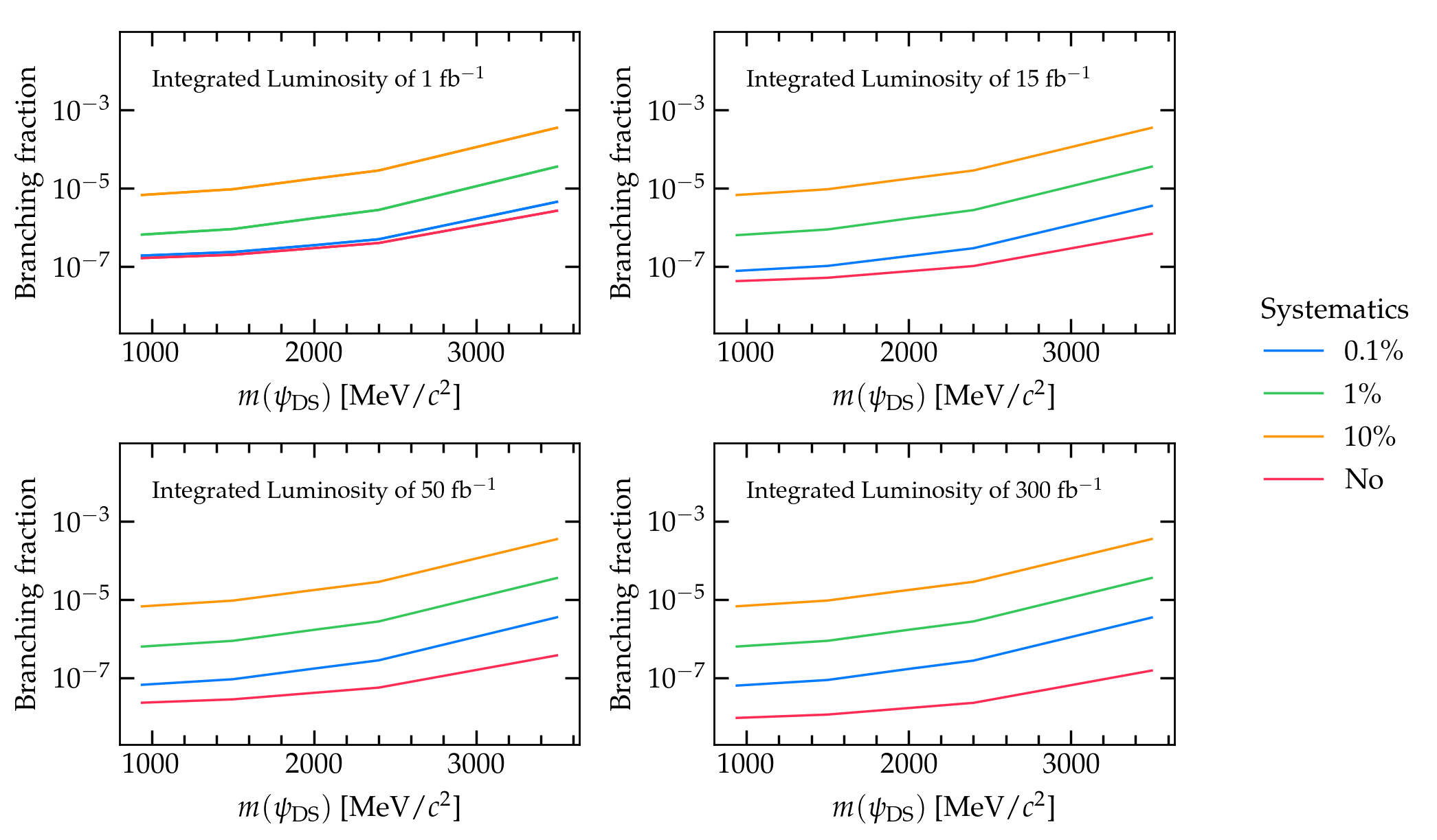}
    \caption{Projected sensitivity in terms of branching fractions excluded at 95\% CL for the $B^0 \rightarrow  \psiDS \Lambda(1520) $ channel. The sensitivity is shown assuming integrated luminosities of 1, 15, 50 and 300 $\textrm{fb}^{-1}$. In each case, different assumptions are made regarding the systematic uncertainty of the background yield, ranging from no uncertainty at all to 10\%.}
    \label{fig:limits_syst_1}
\end{figure}

\begin{figure}[tb]
    \centering
    \includegraphics[width = 0.5\textwidth] {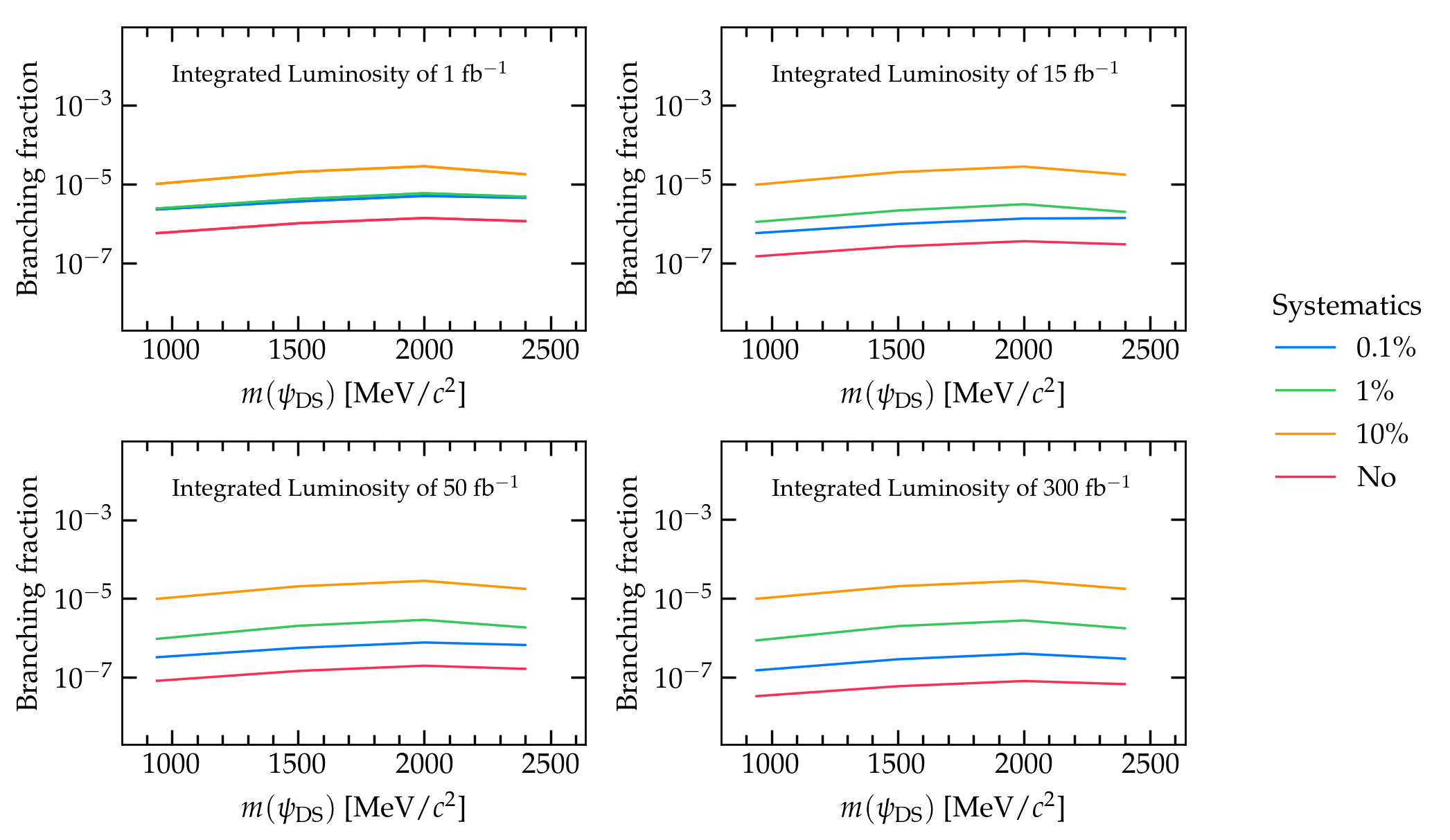}\hfill
    \caption{Same as Fig.~\ref{fig:limits_syst_1}, but now for the $B^+\rightarrow \psiDS \Lambda_c(2595)^+ $ channel.}
    \label{fig:limits_syst_2}
\end{figure}

To provide a more quantitative estimate of the effect of the systematic uncertainty in  the background  yield,  we repeat the  results of Sec.~\ref{sec:sens} for two of our benchmark channels, with the same methodology  but now under different assumptions on the value of this uncertainty.  These can found in Fig.~\ref{fig:limits_syst_1} and Fig.~\ref{fig:limits_syst_2} for the $B^0 \rightarrow  \psiDS \Lambda(1520) $ and $B^+\rightarrow \psiDS \Lambda_c(2595)^+ $ channels, respectively. The main conclusion is that, in a pessimistic scenario, a systematic uncertainty  of $O(10)\%$ would make the search systematically limited already with $\sim 1\,\rm{fb}^{-1}$, compromising the reach of all the region of theoretical interest, with branching fractions $\gtrsim 10^{-5}$, for the channels of interest and any value of $ m_{\psiDS}$.  This region would be completely at reach with uncertainties of $O(1)\%$ or smaller. Finally, bringing the uncertainty down to the level per-mile would increase the systematic limit of the search to $\sim 15\,\rm{fb}^{-1}$.

\bibliographystyle{spphys}
\bibliography{main}

\end{document}